\documentclass[a4paper,11pt]{article}

\usepackage{contribution}

\newcommand{\weblink}[2][]{%
    \ifthenelse{\equal{#1}{}}%
    {\textnormal{\url{#2}}}%
    {\textnormal{\href{#2}{#1}}}%
}

\newcommand{\acknowledgements}[1]{%
  \bigskip\bigskip
  \textsf{\textbf{\Large Acknowledgements}} \\[2ex]
  {#1}
  \bigskip
}

\def\beq{\begin{equation}}
\def\eeq#1{\label{#1}\end{equation}}
\def\eeqn{\end{equation}}

\def\beqa{\begin{eqnarray}}
\def\eeqa#1{\label{#1}\end{eqnarray}}
\def\eeqan{\end{eqnarray}}

\let\bar=\overbar

\def\Dslash{\not{\hbox{\kern-4pt $D$}}}
\def\dslash{\not{\hbox{\kern-2pt $\del$}}}

\def\msb{{\bar{\ssstyle M \kern -1pt S}}}

\newcommand{\contribution}[7][]{%
  \clearpage
  \thispagestyle{plain}
  \ifthenelse{\equal{#1}{}}
  {\hypersetup{pdftitle={#2}}}
  {\hypersetup{pdftitle={#1}}}
  \hypersetup{pdfauthor={{#3} {#4}}}
  {\centering\normalfont\LARGE\bfseries\sffamily #2 \par\nobreak}
  \lhead{}
  \chead{%
    \textit{\footnotesize XIV International Conference on Hadron Spectroscopy
      (\weblink[\textit{hadron2011}]{http://www.hadron2011.de}), 13-17 June 2011, Munich, Germany}%
  }
  \rhead{}
  \bigskip
  \begin{center}
    {#3} {#4}\ifthenelse{\equal{#6}{}}{}{\footnote{\weblink[#6]{mailto:#6}}}
    \ifthenelse{\equal{#7}{}}{}{#7} \\
    \textit{#5}
  \end{center}
  \bigskip
}

\renewcommand{\abstract}[1]{%
  \begin{center}
    \begin{minipage}{0.85\textwidth}
      \begin{footnotesize}
        #1
      \end{footnotesize}
    \end{minipage}
  \end{center}
  \bigskip
}

\begin{document}

%
%
%
%
%
{  


%

\contribution[New results on $\pi^{-}\pi^{0}\pi^{0}$ in comparison to $\pi^{-}\pi^{+}\pi^{-}$ final states]
{\textnormal{Spin-exotic search in the $\rho\pi$ decay channel:}
New results on $\mathbold{\pi^{-}\pi^{0}\pi^{0}}$ in comparison to $\mathbold{\pi^{-}\pi^{+}\pi^{-}}$ final states 
\\
\textnormal{(diffractively produced on proton)}}  
{Frank}{Nerling}  
{Physikalisches Institut, Albert-Ludwigs-Universit\"at Freiburg \\
  79104 Freiburg, GERMANY}  
{nerling@cern.ch}  
{on behalf of the COMPASS Collaboration}  
%

\abstract{%
The COMPASS experiment at CERN SPS features charged particle tracking as well as good coverage by electromagnetic calorimetry, and our data provide an excellent opportunity for simultaneous observation of new states in different decay modes by the same experiment. The existence of the spin-exotic $\pi_1(1600)$ resonance in the $\rho\pi$ decay channel is studied for the first time at COMPASS in both decay modes of the diffractively produced $(3\pi)^{-}$ system: $\pi^{-}p \rightarrow \pi^{-}\pi^{+}\pi^{-}p$ and $\pi^{-}~p \rightarrow \pi^{-}\pi^{0}\pi^{0}~p$. A preliminary partial-wave analysis performed on the 2008 proton target data allows for a first conclusive comparison of both $(3\pi)^{-}$ decay modes not only for main waves but also for small ones. We find the neutral versus charged mode results in excellent agreement with expectations from isospin symmetry. Both, the intensities and the relative phases to well-known resonances, are consistent for the neutral and the charged decay modes of the $(3\pi)^{-}$ system. The status on the search for the spin-exotic $\pi_1(1600)$ resonance produced on a proton target is discussed.
}
%

\section{Introduction}
The existence of exotic states beyond the simple Constituent Quark Model
(CQM) has been speculated about almost since the introduction of colour~\cite{Jaffe:1976,Barnes:1983}. 
So-called hybrid mesons ($q\bar{q}$ states with excited gluonic degree of freedom) and 
glueballs (purely gluonic states without valence quarks) are allowed 
within Quantum Chromodynamics due to the self-coupling of gluons via the colour-charge, 
while they are forbidden within the CQM. 
Even though glueball candidates have been reported by the Crystal Barrel and
the WA102 experiments, the mixing with ordinary isoscalar mesons makes the 
interpretation difficult. Several light hybrids on the other hand are predicted to have exotic 
$J^{PC}$ quantum numbers and are thus promising candidates in the search for resonances beyond the 
CQM. The hybrid candidate lowest in mass is predicted~\cite{Morningstar:2004} to have 
a mass between 1.3 and 2.2\,GeV/$c^2$ and exotic quantum numbers $J^{PC}=1^{-+}$, not 
attainable by ordinary $q\bar{q}$ states.
Several experimentally observed $1^{-+}$ hybrid candidates in the light-quark 
sector have been reported in different decay channels in the past, however, they are all 
still controversially discussed in the community, see e.g.~\cite{MeyerHaarlem2010}.
In particular, the resonant nature of the $\pi_1(1600)$ observed by both E852 and VES in the $\rho\pi$ 
decay channel~\cite{Adams:1998,Khokhlov:2000} in $3\pi$ final states is questioned. 
\begin{figure}[tp!]
    \begin{center}
      \vspace{-0.5cm}
      \includegraphics[clip,trim= 110 120 60 220,width=1.0\linewidth]
      {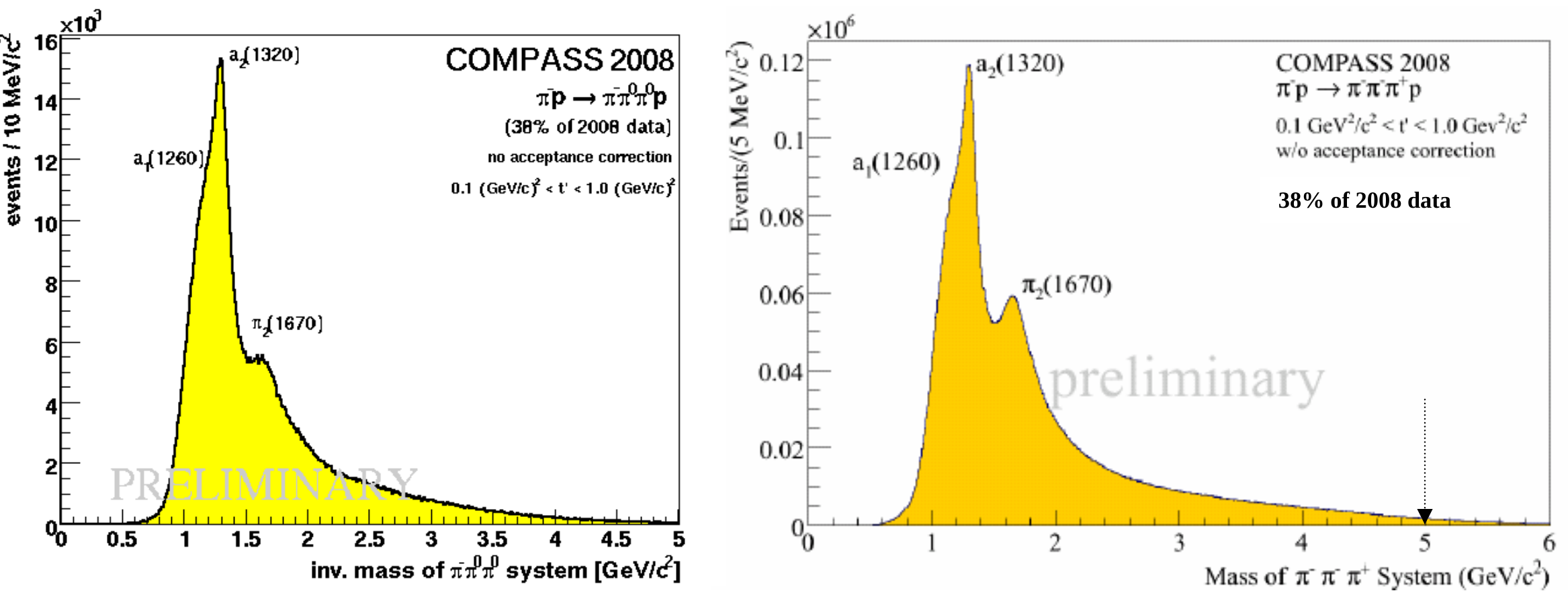}
      \vspace{-0.5cm}
      \caption{Total mass spectrum of the $(3\pi)^{-}$ systems -- neutral \textit{(left)} versus charged decay mode \textit{(right)}.
The spectra look similar as expected, showing both the most prominent, well-known resonances $a_1(1260)$, $a_2(1320)$ and $\pi_2(1670)$.}
      \label{fig:3piMassTot_neutral_charged}
      \vspace{-0.4cm}
    \end{center}
\end{figure}
In later publications, certain conclusions were withdrawn~\cite{Amelin:2005} and re-analyses of the 
$(3\pi)^{-}$ system in two different final states within the same collaboration lead to 
opposite conclusions~\cite{Dzierba:2006}, respectively. One may get a hint at this controversy looking 
at\cite{PDG}. 

After a short pilot run in 2004 (190\,GeV/$c$ $\pi^{-}$ beam, Pb target), we recorded 
high statistics using a 190 GeV/$c$ negative pion beam scattered off a liquid hydrogen (proton) target. 
A similar amount of data with 190\,GeV/$c$ positive hadron beams has been taken in 2009, as well
as some data (negative beam) with nuclear targets. 
As a first input to the puzzle, COMPASS observed a significant $J^{PC}$ spin-exotic signal 
in the 2004 data (in three charged pion final states) consistent with the disputed $\pi_1(1600)$ 
that was accepted for publication last year~\cite{Alekseev:2009a}. The proton and nuclear target data 
taken in 2008/09 will enable COMPASS to further clarify the situation.

\section{New results of $\mathbold{\pi^{-}\pi^{0}\pi^{0}}$ in comparison to $\mathbold{\pi^{-}\pi^{+}\pi^{-}}$ final states} 
\label{sec.PWAresults}
The invariant mass of the $(3\,\pi)^{-}$ system is shown for the neutral and the charged $\rho\pi$ 
decay modes in Fig.\,\ref{fig:3piMassTot_neutral_charged}. About half of the 2008 data with negative 
pion beam of 190\,GeV/$c$ have been analysed so far in the high momentum transfer $t'$ region of 
$0.1\,{\rm GeV}^2/c^2 < t' < 1.0\,{\rm GeV}^2/c^2$. Due to the different detection 
efficiencies obtained for neutral and charged particles, this translates to $\sim$1\,M events (neutral 
mode) and $\sim$24\,M event (charged mode), respectively, being roughly in the range as expected in general.

The new mass-independent PWA results for neutral and charged mode data presented in this paper, are
normalised using the $a_2(1320)$ as a standard candle as shown in Fig.\,\ref{fig:isospinSymmMainWaves} (top, left)
in order to compensate for the different detection efficiencies. This makes the fitted intensities for individual 
partial waves comparable between neutral and charged modes. 
With respect to \cite{nerling:2009}, the wave-set has been extended from 42 to 53 waves, cf. also~\cite{haas:2011} 
giving further details on the analysis of the $\pi^{-}\pi^{+}\pi^{-}$ final states likewise, and the data analysed 
for the neutral mode has been increased by about a factor of five, allowing for a first conclusive comparison also 
for small waves. A detailed description of the applied PWA method can be found in~\cite{nerling:2009,haas:2011} and 
references therein.

Even though the neutral mode data have not yet been corrected for acceptance (which is rather flat for the charged mode), 
our data is in good agreement with expectations from isospin coupling considerations. 
If an isospin 1 resonance decays into $\rho\pi$, similar intensities are expected for neutral and charged mode, 
whereas decays into $f_2\pi$ should show a suppression factor of two for the neutral w.r.t. to the charged mode data, 
simply due to the Clebsch-Gordan coefficients determining the different isospin coupling for the different underlying 
isobar structure, i.e. decays into an isovector versus an isoscalar.
\begin{figure}[tp!]
  \begin{minipage}[h]{.32\textwidth}
    \begin{center}
         \vspace{-0.3cm}
     \includegraphics[clip,trim= 5 20 23 10, width=1.0\linewidth, angle=90]{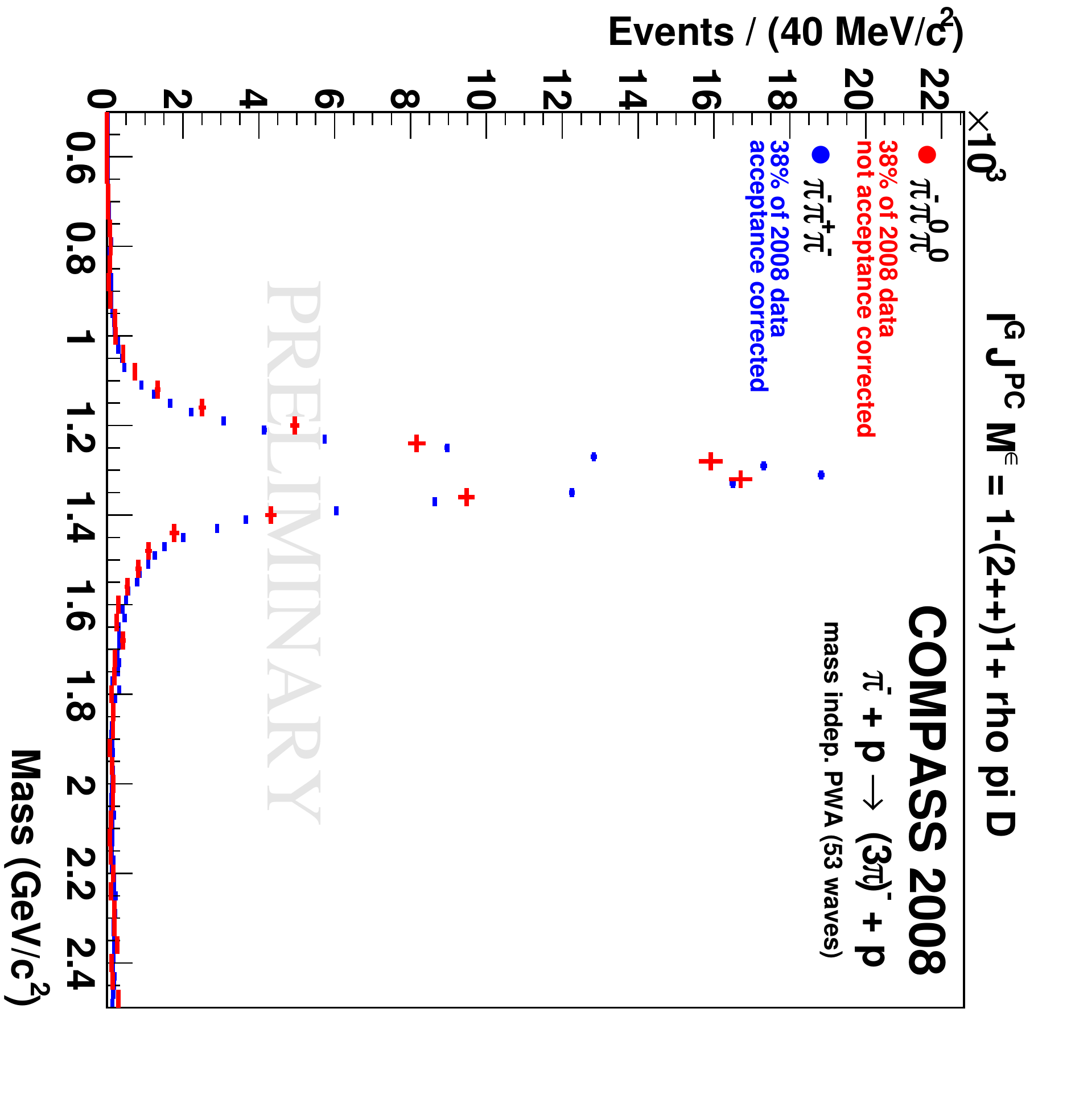}
    \end{center}
  \end{minipage}
  \hfill
  \begin{minipage}[h]{.32\textwidth}
    \begin{center}
      \vspace{-0.3cm}
     \includegraphics[clip,trim= 5 28 23 8, width=1.0\linewidth, angle=90]{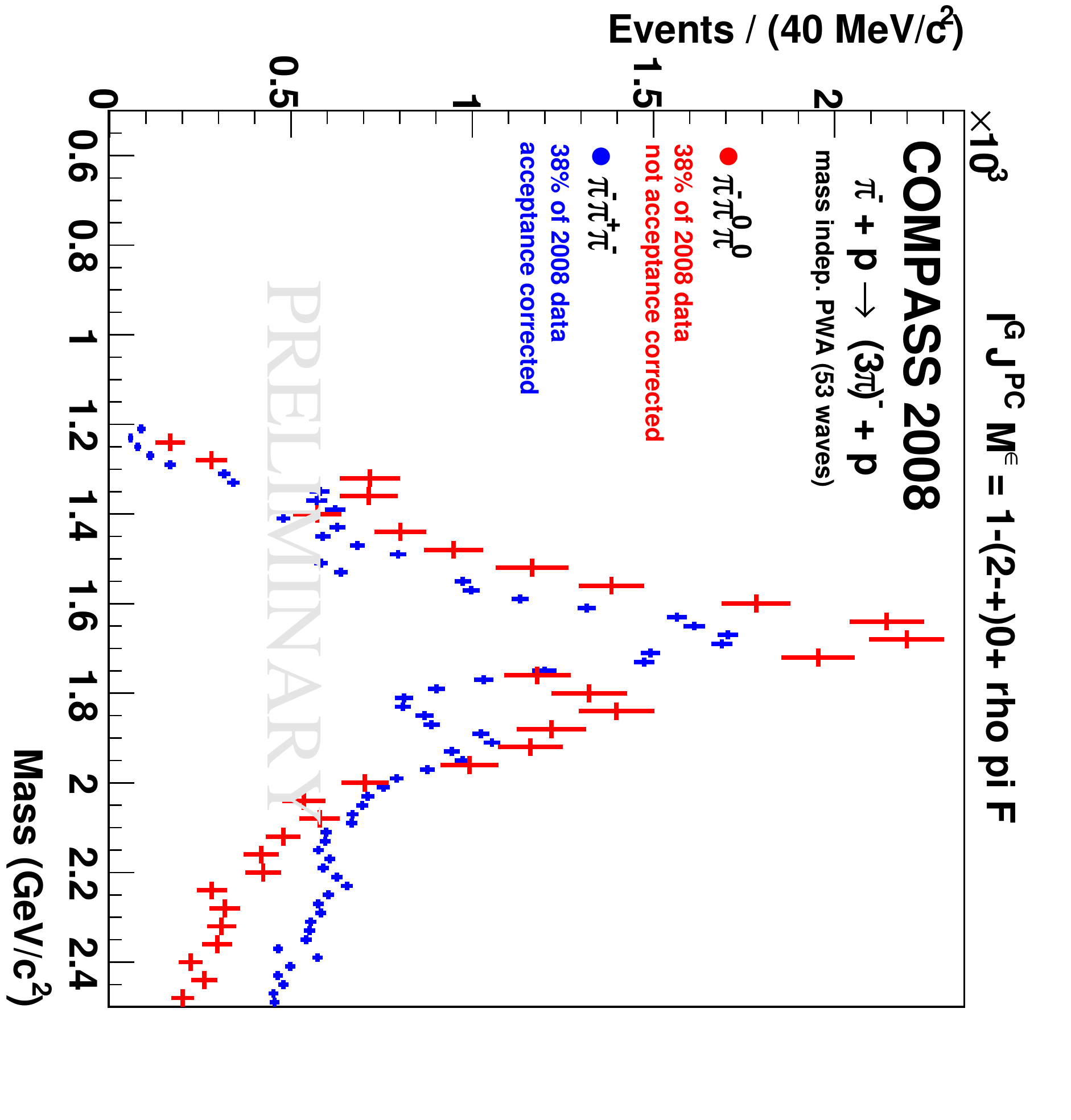}
    \end{center}
  \end{minipage}
  \begin{minipage}[h]{.32\textwidth}
    \begin{center}
      \vspace{-0.3cm}
     \includegraphics[clip,trim= 5 0 23 0, width=1.0\linewidth, angle=90]{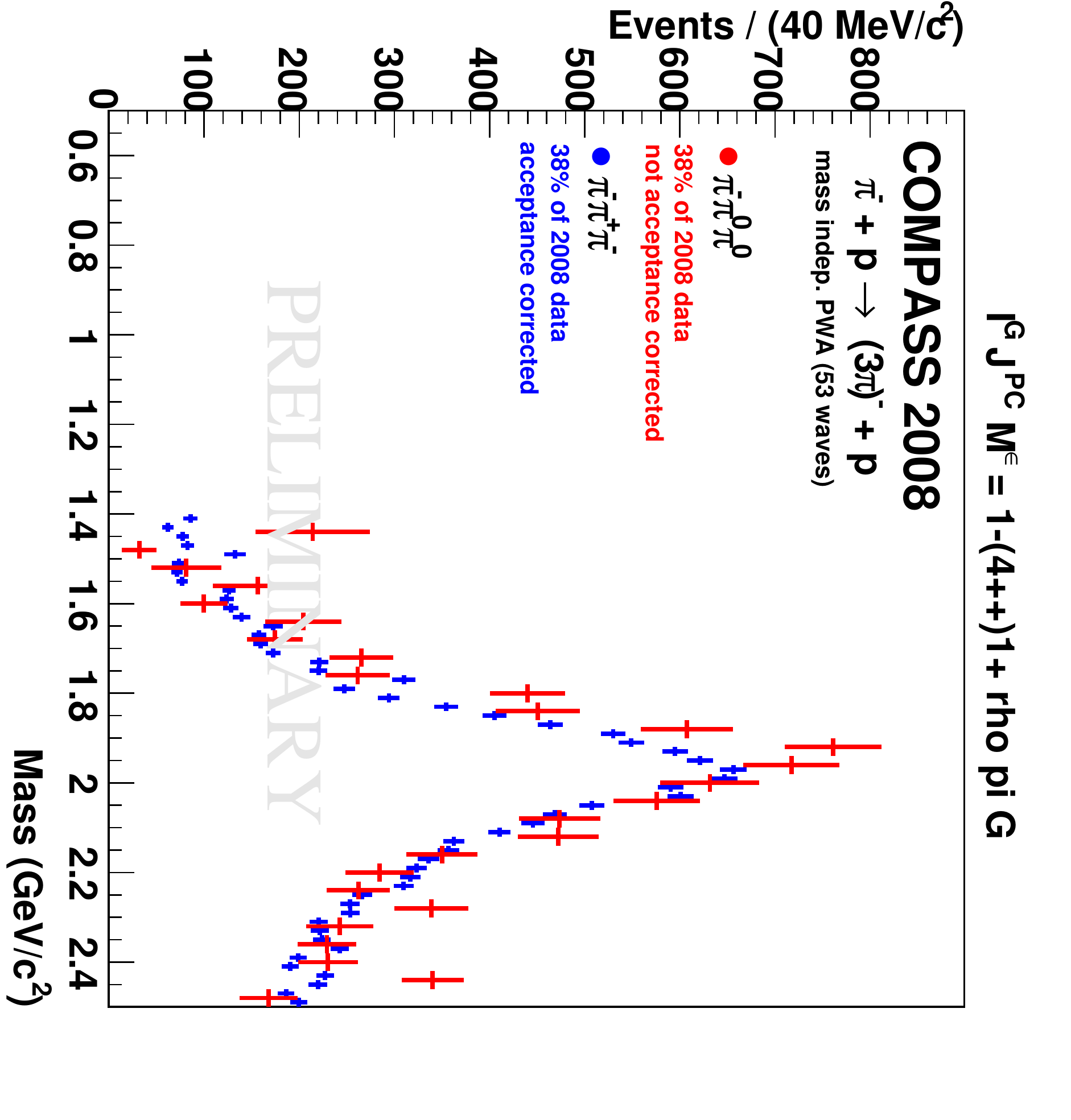}
    \end{center}
  \end{minipage}
  \begin{minipage}[h]{.32\textwidth}
    \begin{center}
         \vspace{-0.3cm}
     \includegraphics[clip,trim= 5 20 23 10, width=1.0\linewidth, angle=90]{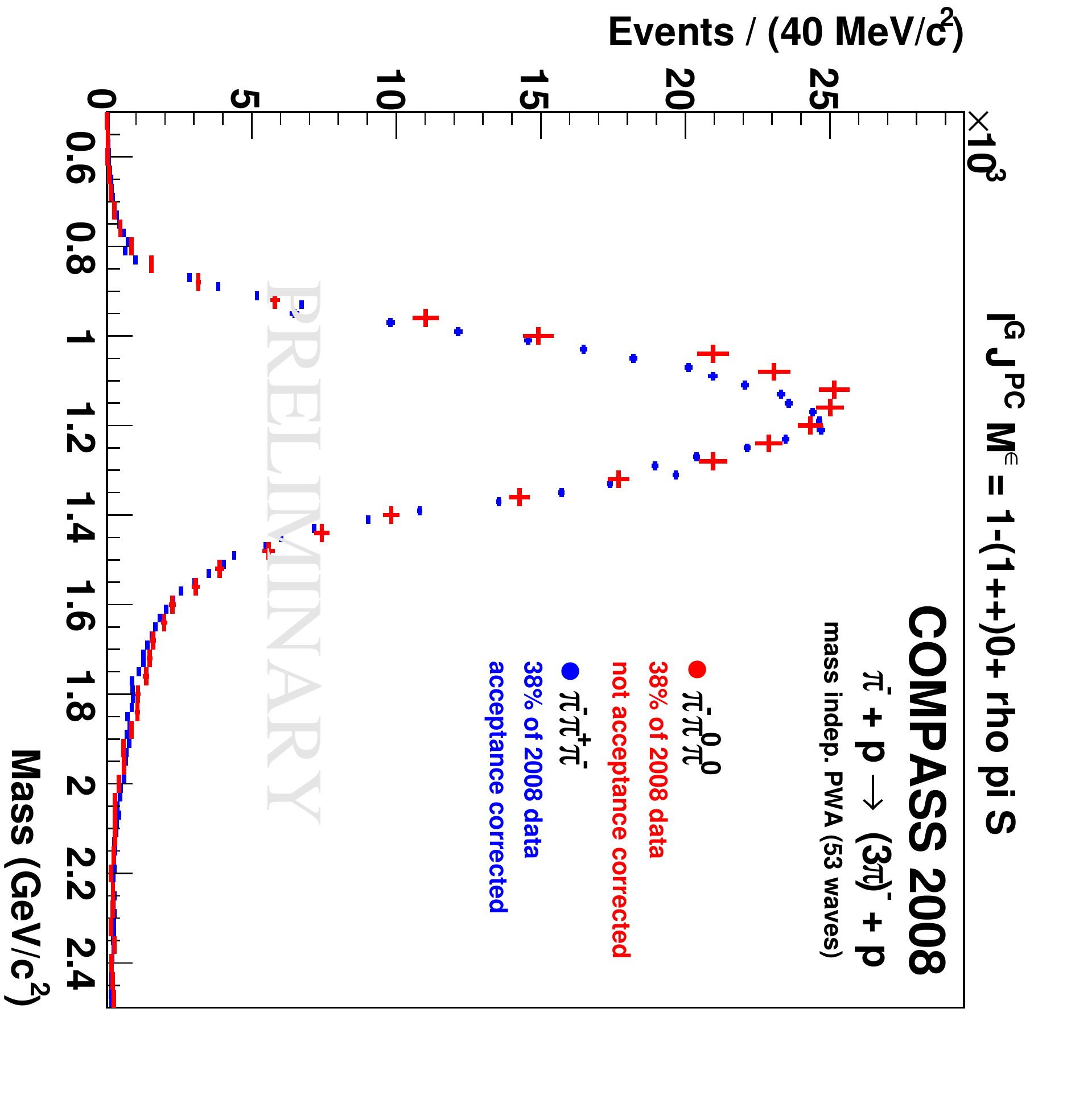}
    \end{center}
  \end{minipage}
  \hfill
  \begin{minipage}[h]{.32\textwidth}
    \begin{center}
     \vspace{-0.3cm}
     \includegraphics[clip,trim= 5 52 23 -16, width=1.0\linewidth, angle=90]{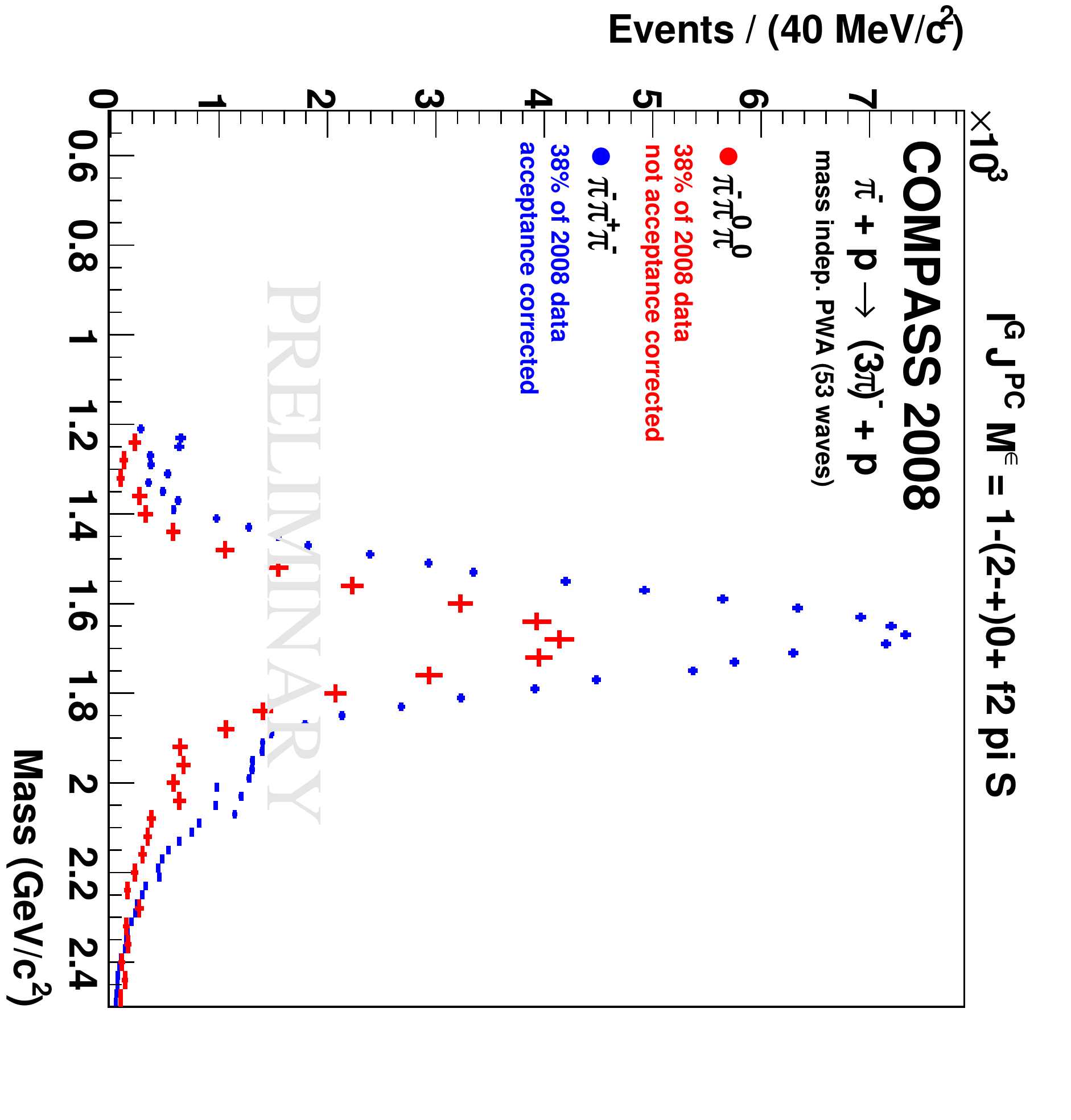}
    \end{center}
  \end{minipage}
  \hfill
  \begin{minipage}[h]{.32\textwidth}
    \begin{center}
      \vspace{-0.3cm}
     \includegraphics[clip,trim= 5 48 23 -12, width=1.0\linewidth, angle=90]{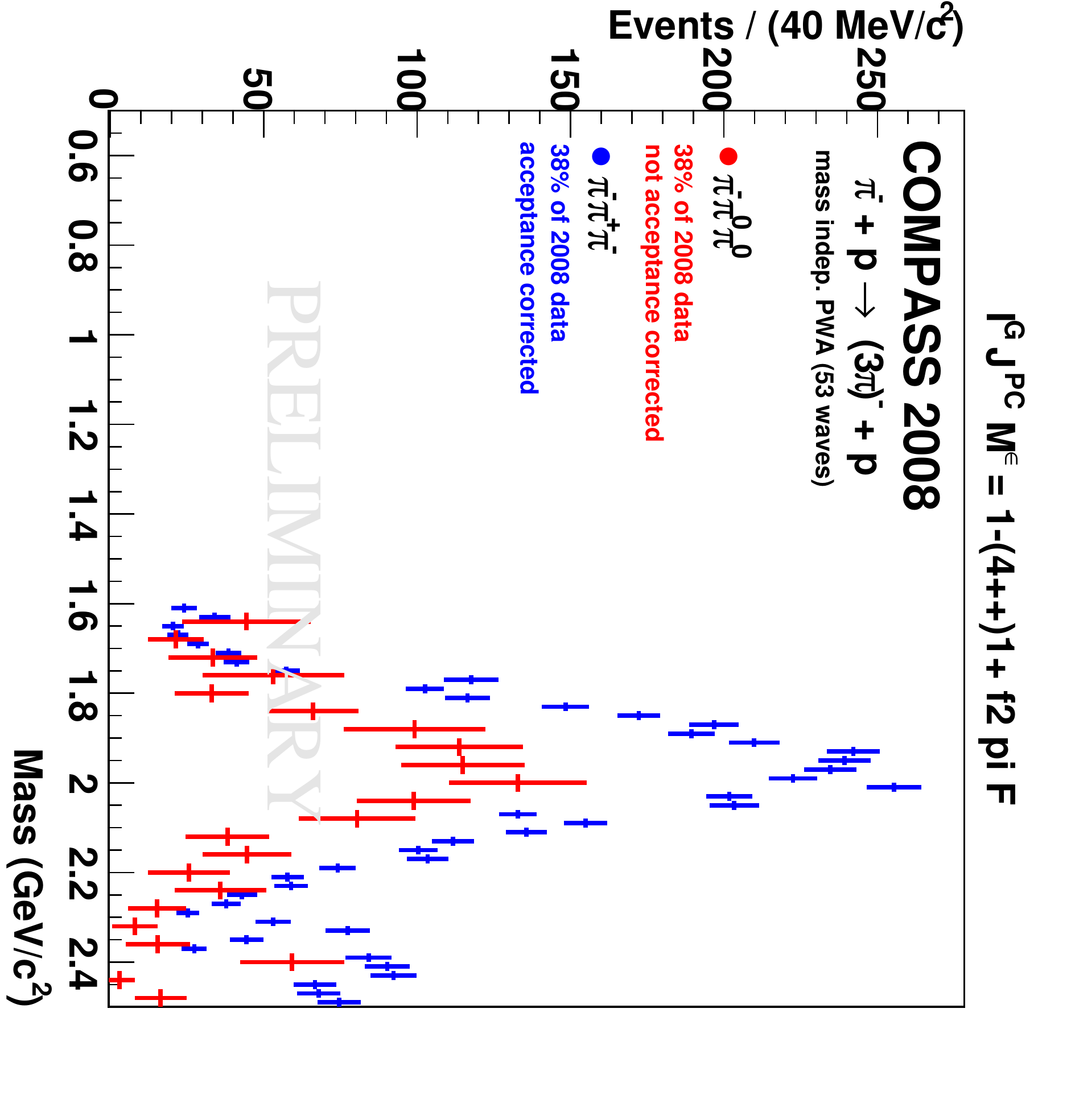}
    \end{center}
  \end{minipage}
     \vspace{-0.4cm}
    \begin{center}
     \caption{Comparison of PWA mass-independent fit result for neutral versus charged mode -- exemplary main and small waves: {\it (Top, left)} $a_2(1320)$ used for normalisation, {\it (top \& bottom, centre)} $\pi_2(1670)$, {\it (top \& bottom, right)} $a_4(2040)$, and {\it (bottom, left)} $a_1(1260)$, 
respectively {\it (red = neutral, blue = charged)}. For discussion see text.}
       \label{fig:isospinSymmMainWaves}
     \end{center}
     \vspace{-0.3cm}
\end{figure}
This is shown for some exemplary main and small waves in Fig.\,\ref{fig:isospinSymmMainWaves}. The $a_1(1260)$ decaying into 
$\rho\pi$ is observed with same width and intensity for both modes, similarly for the $\pi_1(1670)$ and $a_4(2040)$ decays
into $\rho\pi$, whereas a suppression factor of about two is observed for the neutral mode intensities as compared to the 
charged mode data for the resonances decaying into $f_2\pi$ --- as expected.
That our data follows the expectation from isospin symmetry to a large extent throughout the whole wave-set is shown 
by Fig.\,\ref{fig:isospinSymmSpinTotals-53w}, depicting the intensity sums of all $\rho\pi$ and $f_2\pi$ partial waves, respectively.

\begin{figure}[tp!]
  \begin{minipage}[h]{.49\textwidth}
    \begin{center}
\vspace{-0.7cm}
     \includegraphics[clip,trim= 3 4 22 5, width=0.85\linewidth, angle=90]{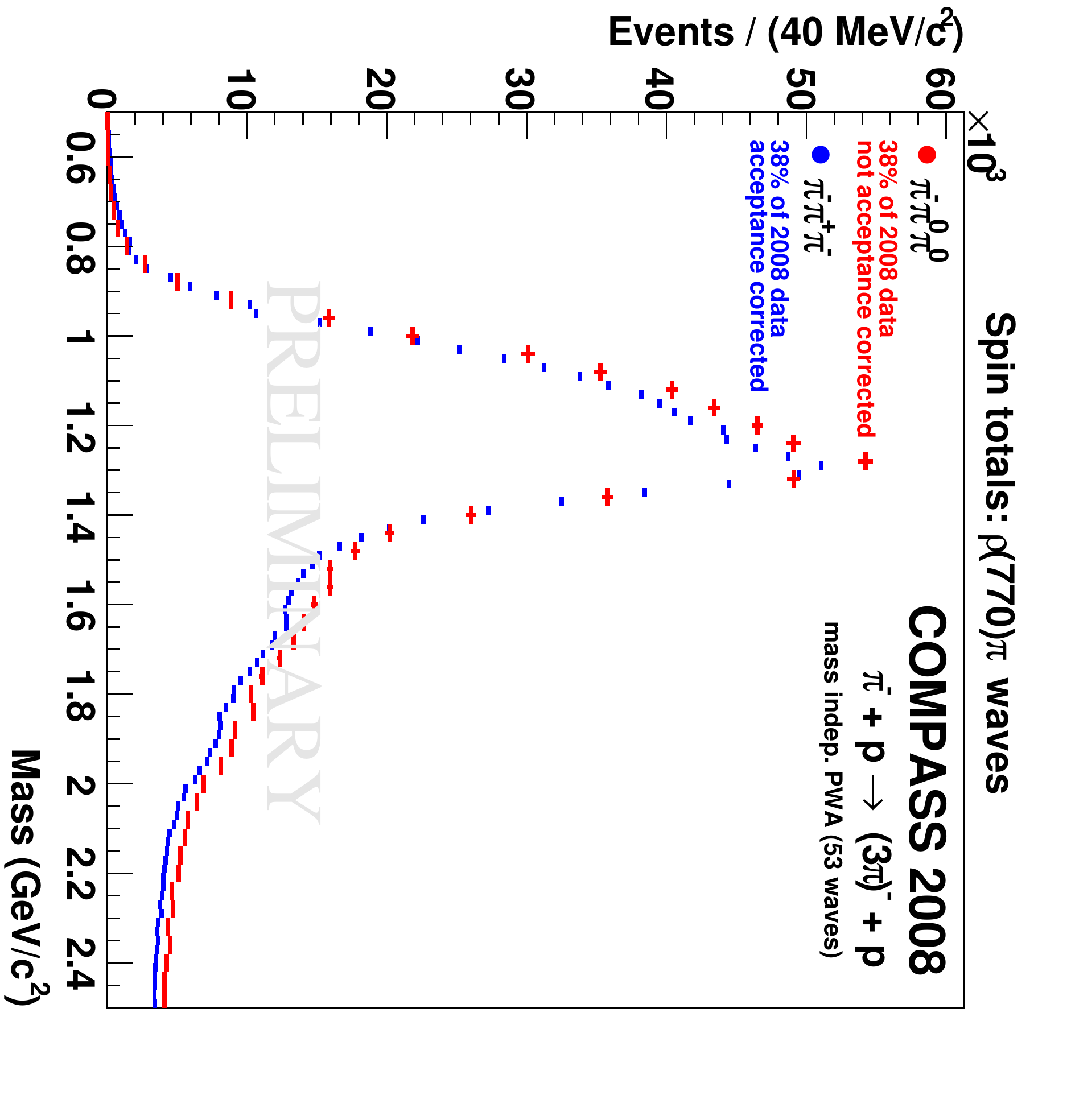}
    \end{center}
  \end{minipage}
  \hfill
  \begin{minipage}[h]{.49\textwidth}
    \begin{center}
\vspace{-0.7cm}
     \includegraphics[clip,trim= 3 4 22 5,angle=90, width=0.9\linewidth, angle=0]{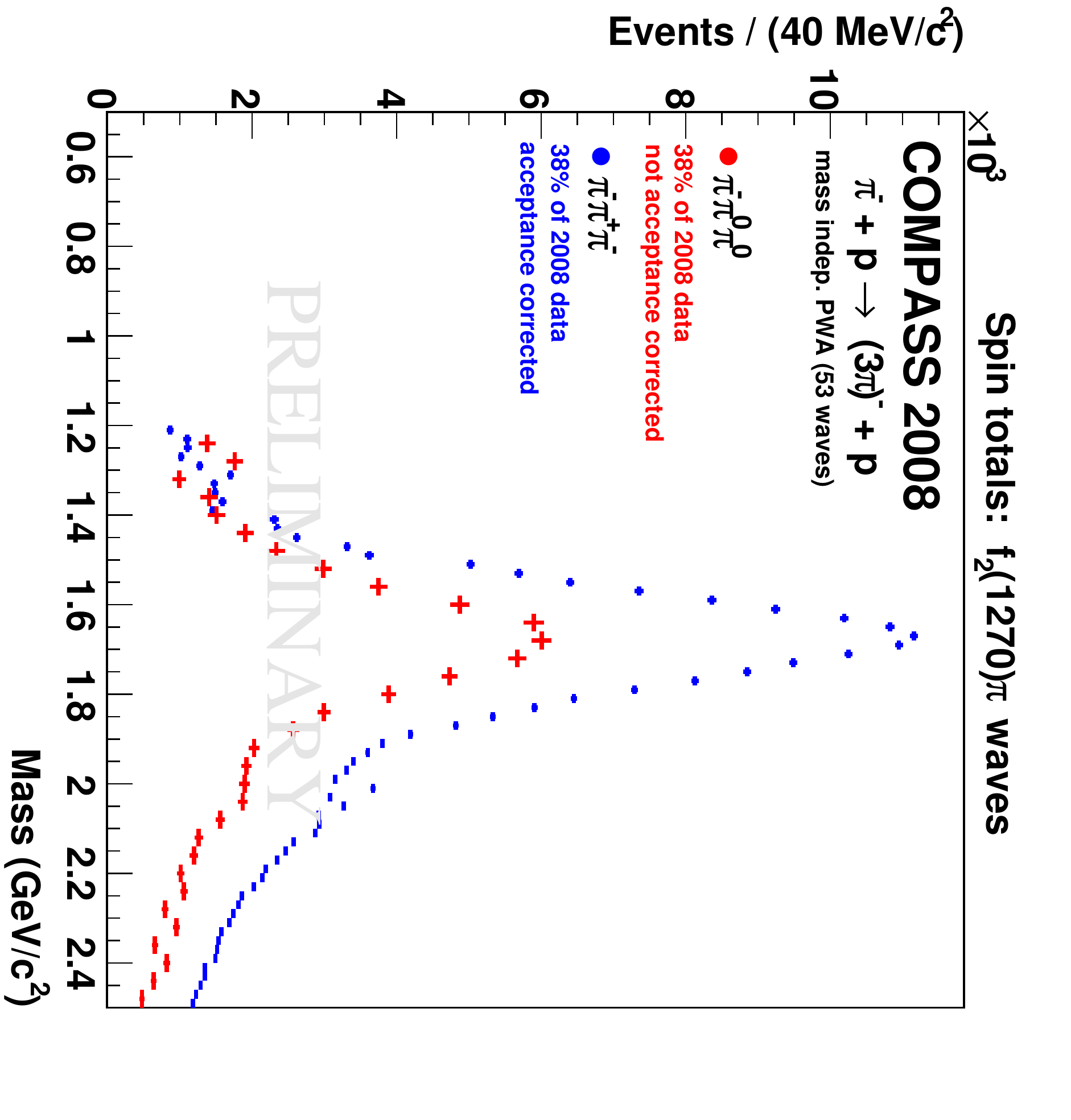}
    \end{center}
  \end{minipage}
      \vspace{-0.2cm}
     \caption{Intensity sums of all partial waves decaying into $\rho\pi$ {\it (left)} and $f_2\pi$ {\it (right)} intermediate states included in the fit are in good agreement with expectations from isospin symmetry. For discussion see text.}
       \label{fig:isospinSymmSpinTotals-53w}
\end{figure}
%
This simple isospin-symmetry holds only presumed the branchings are entirely determined by the isospin 
Clebsch-Gordan coefficients, which is not true in general for $f_{0,2}\pi$ decays, as Bose-Symmetrisation 
with the other $\pi^{-}$ and $\pi^{0}$, respectively, is required and might change the observed ratio of intensities due to interference 
effects. There is no such effect, however, for $\rho\pi$ decays since whatever the effect might be, it is the 
same for both decay modes, and therefore cancels out. 
We checked by calculation using the wave function, that there is no such effect for 
$\pi_2(1670)$ decaying into $f_2\pi$, for which the pure suppression factor of two is 
indeed expected, as given in Table~\ref{tab:BRpdg}.  
\begin{table}[bp!]
  \centering
  \begin{tabular}[]{llll} \hline
    BR = N($\pi^-\pi^0\pi^0 $  )/N($\pi^-\pi^-\pi^+ $  ) -- calculated from isobar model amplitudes          \\ \hline
    BR( $ 0^{-+} f_0(980) \pi$~$ S $) =  0.44 (at 1.8 GeV)            \\
    BR( $ 1^{++} (\pi\pi)_s \pi $~$P $) = 0.80 (at 1.3 GeV)            \\
    BR( $ 2^{-+} f_2(1270) \pi $~$ S $) =  0.50 (at 1.67 GeV)         \\    
    \hline 
  \end{tabular}
  \caption{Isospin symmetry and final state Bose-Symmetrisation: Calculation of the relative branching ratios (BR) of neutral to charged mode 
for decays via different isobars. The isospin Clebsch-Gordan coefficients have been applied inside the PWA normalisation 
integral calculator to calculate the BR for the different partial waves.}
  \label{tab:BRpdg}
\end{table} 
Depending on the overlap of the isobars on the Dalitz plot, interference effects might change the factor of two, as 
for example in case of the $\pi(1800)$ decay into $f_0(980)\pi$, for which we expect to find an enlarged suppression 
factor (Tab.\,\ref{tab:BRpdg}) in good agreement with our data, see Fig.\,\ref{fig:phases_a1_a2__a1_pi2-53w} (top, centre). 
Apart of the fitted intensities it is conclusive to look at the phase difference of a possible resonance
with respect to a well-know one. 
If the candidate is indeed a resonance, connected with the reference one (but not phase-locked), it should manifest in a 
clean phase motion between them, as it is shown for main and small waves in Fig.\,\ref{fig:phases_a1_a2__a1_pi2-53w},
using the prominent $a_1(1260)$ as reference.
For the $a_2(1320)$, we find a rapid phase motion just in the range between the maxima of both objects, consistently 
coinciding for the neutral and the charged mode data. For the $\pi_2(1670)$ and the $a_4(2040)$, we observe a clean, 
rapid phase motion as well, again consistently coinciding for both modes. As they are more separated in mass 
from the reference, they are resonating against the tail of the $a_1(1260)$ resulting in observed phase motions limited 
to the mass range (about 1.7-1.9 and 1.7-2.0\,GeV/$c^2$, respectively) of the given resonance under study.  
\begin{figure}[tp!]
  \begin{minipage}[h]{.32\textwidth}
    \begin{center}
         \vspace{-0.6cm}
     \includegraphics[clip,trim= 0 0 0 0, width=1.0\linewidth, angle=90]{Plots/h14.pdf}
    \end{center}
  \end{minipage}
  \hfill
  \begin{minipage}[h]{.32\textwidth}
    \begin{center}
      \vspace{-0.6cm}
     \includegraphics[clip,trim= 0 0 0 0, width=1.0\linewidth, angle=90]{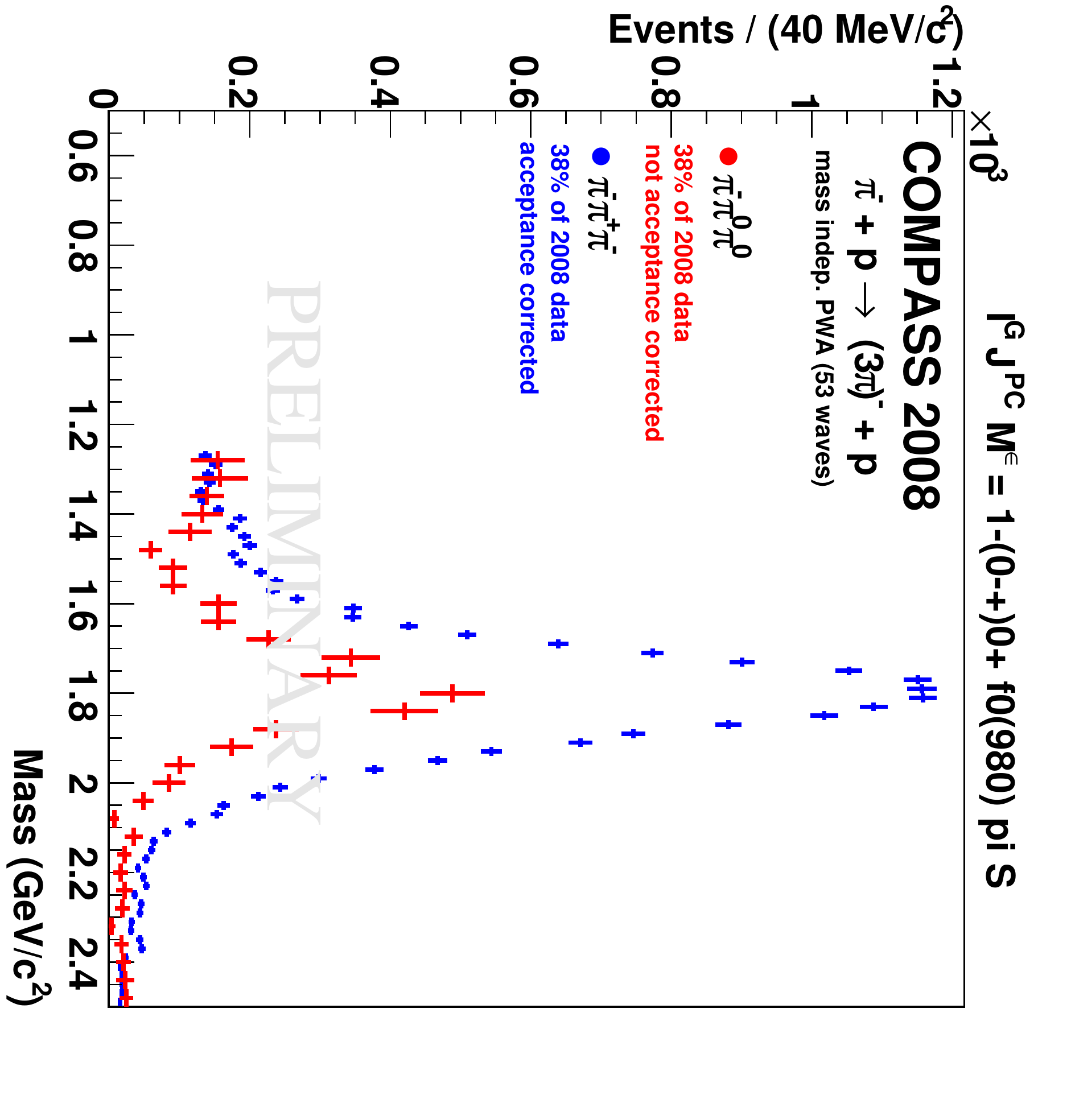}
    \end{center}
  \end{minipage}
  \hfill
  \begin{minipage}[h]{.32\textwidth}
    \begin{center}
      \vspace{-0.6cm}
     \includegraphics[clip,trim= 0 0 0 0, width=1.0\linewidth, angle=90]{Plots/h32mod.pdf}
    \end{center}
  \end{minipage}
  \begin{minipage}[h]{.32\textwidth}
    \begin{center}
           \vspace{-0.3cm}
     \includegraphics[clip,trim= 0 0 0 0, width=1.0\linewidth, angle=90]{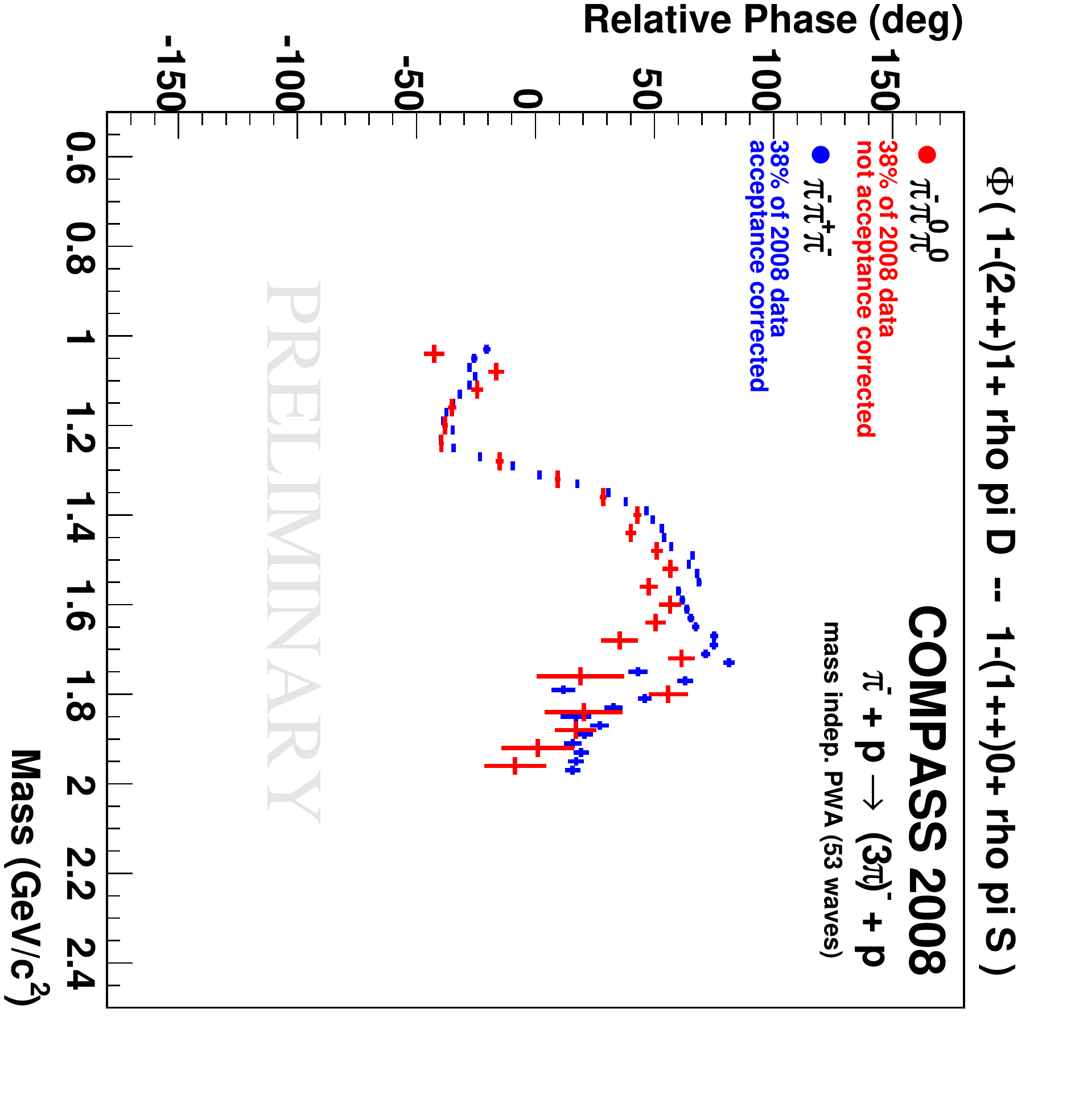}
    \end{center}
  \end{minipage}
  \hfill
  \begin{minipage}[h]{.32\textwidth}
    \begin{center}
      \vspace{-0.3cm}
     \includegraphics[clip,trim= 0 0 0 0, width=1.0\linewidth, angle=90]{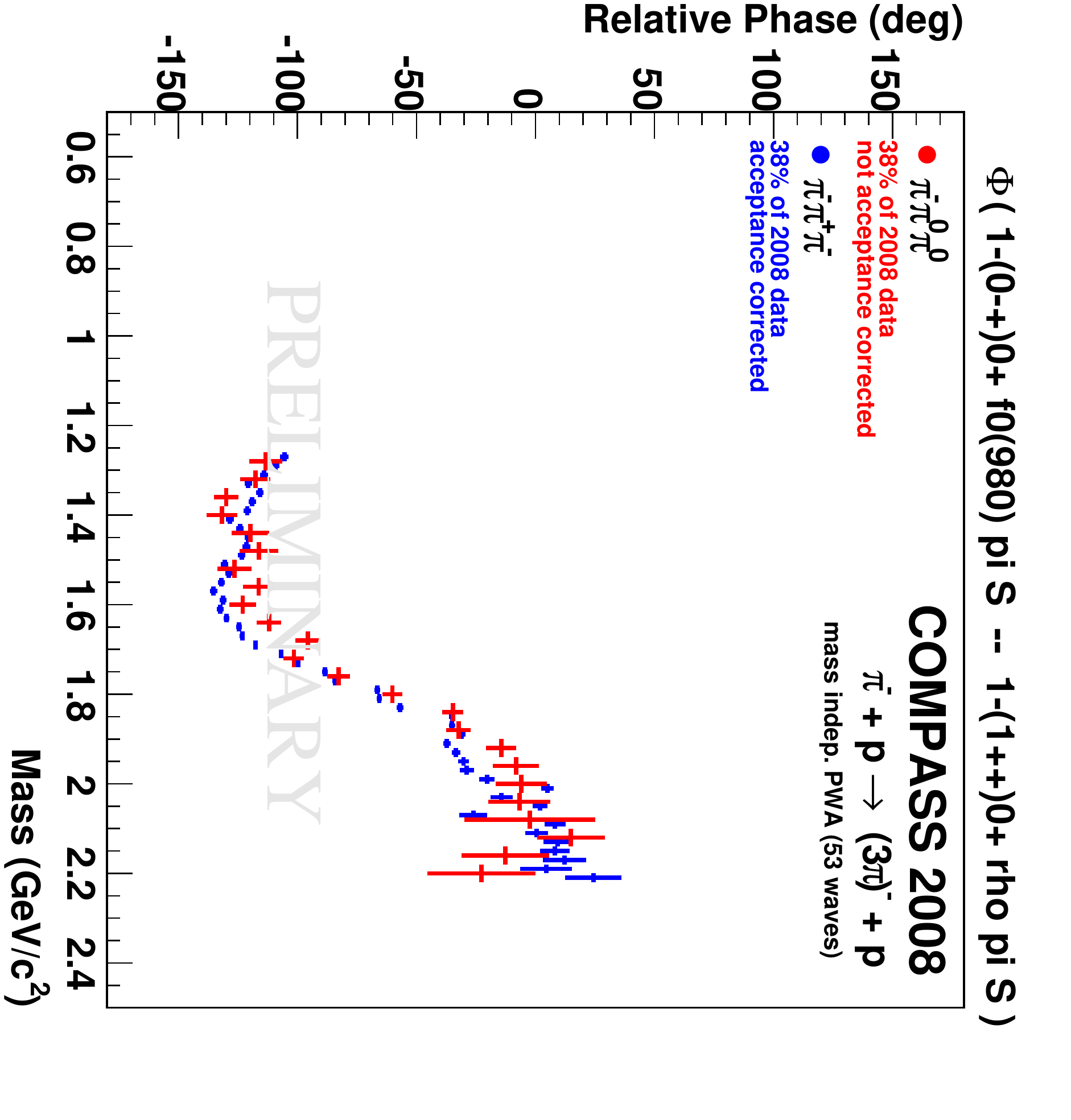}
    \end{center}
  \end{minipage}
  \hfill
  \begin{minipage}[h]{.32\textwidth}
    \begin{center}
      \vspace{-0.3cm}
      \includegraphics[clip,trim= 0 0 0 0, width=1.0\linewidth, angle=90]{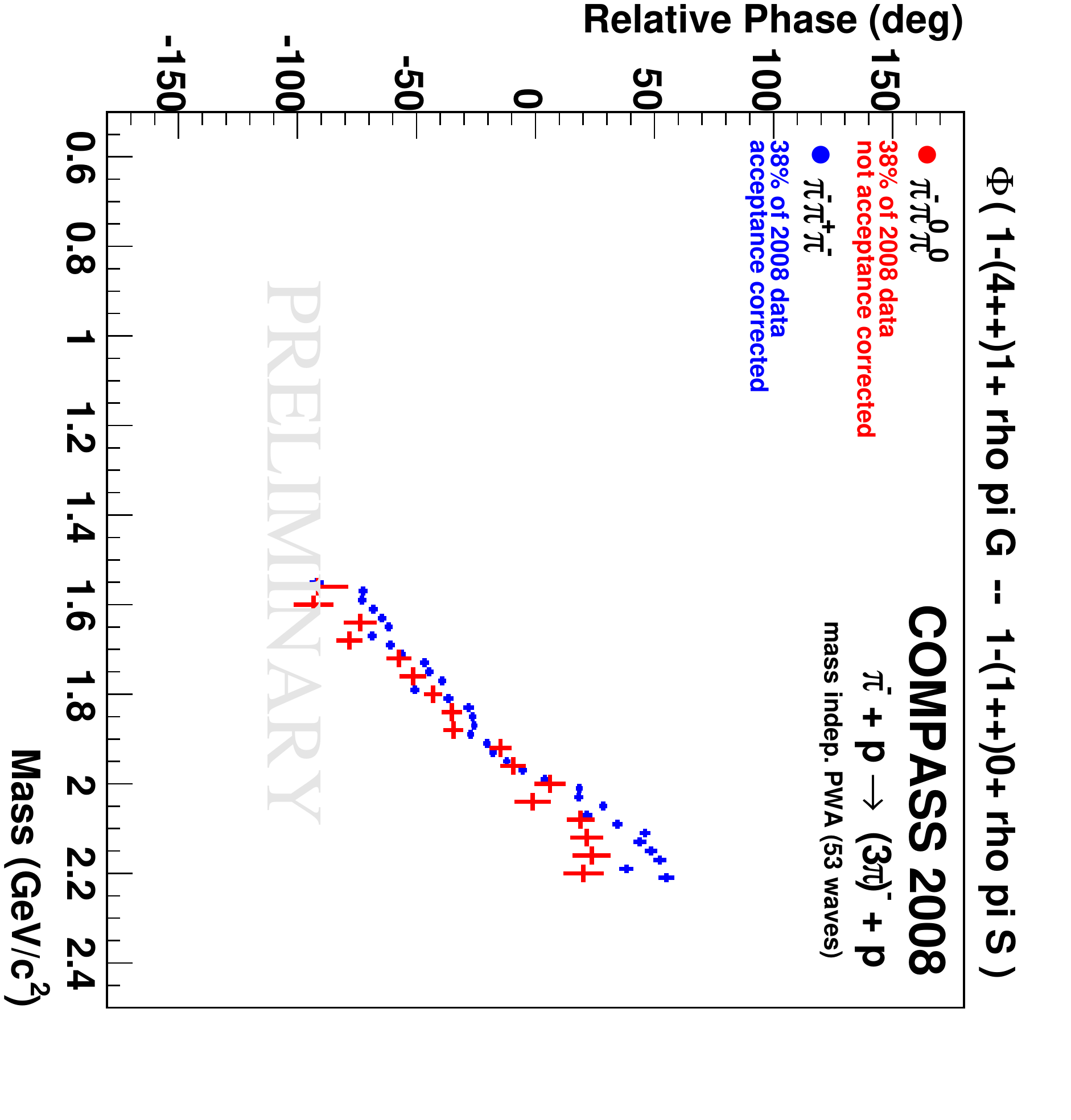}
    \end{center}
  \end{minipage}
  \begin{center}
    \caption{Relative phase difference $\Phi$ for main and small waves with respect to the 
prominent $a_1(1260)$ (see Fig.\ref{fig:isospinSymmMainWaves}, {\it left, bottom}): {\it (Left)}
$a_2(1320) \rightarrow \rho\pi$, {\it (centre)} $\pi(1800) \rightarrow f_0(980)\pi$, and {\it (right)}
$a_4(2040)\rightarrow \rho\pi$, respectively {\it (red = neutral, blue = charged)}. 
The well-known $a_2(1320)$ resonance as well as the smaller, less prominent ones show a clean, rapid phase 
motion relatively to the $a_1(1260)$. Not only the intensities but also the phases are consistent for both, 
neutral and charged mode, for discussion see text.}
    \label{fig:phases_a1_a2__a1_pi2-53w}
  \end{center}
  \vspace{-0.3cm}
\end{figure}

\section{Conclusions \& summary}
The hadron data taken in 2008/09 will allow COMPASS to contribute solving the puzzle of light 
spin-exotic mesons. The high statistics and the possibility of detecting final states involving 
neutral particles allows for simultaneous observation and confirmation of new states in different final states 
by the same experiment. The new results presented on the $\rho\pi$ decay channel in both, neutral and charged decay 
modes of the $(3\pi)^{-}$ system, appear very consistent and solid not only for main but also for small waves.
There is presently no contradiction between both analysis results. In particular the coinciding relative phases 
of various resonances confirm already now the excellent potential to conclude on the existence of the 
spin-exotic $\pi_1(1600)$ resonance in the $\rho\pi$ decay channel, simultaneously observed in two 
different final states with the same experiment. 
  
\acknowledgements{
This work is supported by the BMBF (Germany), especially via the ``Nutzungsinitiative CERN''.
}


%

}  



\begin{thebibliography}{99}
\bibitem{Jaffe:1976} R.~Jaffe and K.~Johnsons, {\it Phys. Lett. B} {\bf 60} (1976) {201}.
\bibitem{Barnes:1983} T.~Barnes {\it et al.}, {\it Nucl. Phys. B} {\bf 224} (1983) {241}.
\bibitem{Morningstar:2004} K.J.~Juge, J.~Kuti, C.~Morningstar, {\it AIP Conf. Proc.} {\bf 688} (2004) {193}.
\bibitem{MeyerHaarlem2010} C.A.~Meyer and Y.Van.~Haarlem, {\it Phys. Rev. C} {\bf 82} (2010) {025208}.
\bibitem{Adams:1998} G.~S.~Adams {\it et al.}, {\it Phys. Rev. Lett.} {\bf 81}, (1998) 5760.
\bibitem{Khokhlov:2000} Y.~Khokhlov, {\it Nucl. Phys.} {\bf A663} (2000) 596.
\bibitem{Amelin:2005} D.~V.~Amelin {\it et al.}, {\it Phys. Atom. Nucl.} {\bf 68} (2005) 359.
\bibitem{Dzierba:2006} A.R.~Dzierba {\it et al.}, {\it Phys. Rev. D} {\bf 73} (2006) {072001}.
\bibitem{PDG} K.~Nakamura {\it et al.} (Particle Data Group), {J. Phys. G} {\bf 37} (2010) {075021}.
\bibitem{Alekseev:2009a} M.~Alekseev {\it et al.}, COMPASS collaboration,   
{\it Phys. Rev. Lett}, {\bf 104} (2010) {241803}.
\bibitem{nerling:2009} F.~Nerling, {\it AIP Conf. Proc.} {\bf 1257} (2010) 286; arXiv:1007.2951[hep-ex].
\bibitem{haas:2011} F.~Haas, {\it These proceedings} (2011).
\end{thebibliography}
\end{document}